\documentclass[aps,prd,10pt,superscriptaddress,preprintnumbers,nofootinbib,twocolumn, floatfix]{revtex4-2}

\usepackage[utf8]{inputenc}
\usepackage{uniinput}

\usepackage{amsmath}
\usepackage{mathtools}
\usepackage{amssymb}
\usepackage{booktabs}
\usepackage{comment}

\usepackage{multirow}
\usepackage{tabularray}

\usepackage{graphicx}
\usepackage{soul}

\usepackage{ragged2e}
\makeatletter
\long\def\@makecaption#1#2{%
  \vskip\abovecaptionskip
  {\footnotesize
   \sbox\@tempboxa{#1\space #2}%
   \ifdim \wd\@tempboxa > \hsize
     \noindent \justifying {\bfseries #1.}\space #2\par
   \else
     \hbox to \hsize{\hfil\box\@tempboxa\hfil}%
   \fi
  }%
  \vskip\belowcaptionskip
}
\makeatother

\usepackage{bm}

\usepackage[utf8]{inputenc}

\usepackage{tikz}
\usepackage{pgfplots}
\pgfplotsset{compat=1.18}

\usepackage{xcolor}

\definecolor{dark-green}{rgb}{0.1,0.4,0}
\definecolor{NiceBlue}{rgb}{0.30196,0.55294,0.57647}

\usepackage{hyperref}
\hypersetup{
	colorlinks=true,
	allcolors=magenta
}

\usepackage{cleveref}

\crefname{footnote}{footnote}{footnotes}
\Crefname{footnote}{Footnote}{Footnotes}

\crefformat{footnote}{#2#1#3}
\Crefformat{footnote}{#2#1#3}

\newcommand{\bea}{\begin{eqnarray}}
\newcommand{\eea}{\end{eqnarray}}

\def\Im{ \hbox{\rm Im}}

\def\Re{\hbox {\rm Re}}
\def\Im{\hbox {\rm Im}}

\begin{document}

\title{Dark bubbles, dark dimensions and fat gravitons}
	
\author{Ulf Danielsson~}
\email{ulf.danielsson@physics.uu.se}
\affiliation{Institutionen för fysik och astronomi,
	Uppsala Universitet, Box 803, SE-751 08 Uppsala, Sweden}
\author{Suvendu Giri~}
\email{suvendu.giri@physics.uu.se}
\affiliation{Institutionen för fysik och astronomi,
	Uppsala Universitet, Box 803, SE-751 08 Uppsala, Sweden}
	
\preprint{UUITP-14/26}

\begin{abstract}
\noindent
The dark bubble model explains the existence of a positive cosmological constant by making explicit use of the instabilities underlying the de Sitter swampland conjectures to make the accelerated expansion of the universe inevitable. A distinctive consequence of the construction is a unique hierarchy connecting cosmological, gravitational, string, and higher-dimensional scales. In particular, the model naturally predicts the existence of a dark dimension of micron size, an idea that has been argued for on independent grounds in the literature. The same framework also predicts a weakening of gravity at distances of order the dark-dimension scale, leading to a fading of the gravitational force at micron distances. We argue that the dark bubble therefore provides a concrete realization of both the dark dimension proposal and Sundrum's fat graviton scenario, in which gravity effectively ceases to probe shorter distances. Additional predictions include a string scale of order tens of TeV and a measurable positive spatial curvature of the universe. We review these key aspects of the model, discuss their implications for gravity and cosmology, and highlight its key predictions.
\end{abstract}

\maketitle
\tableofcontents

\allowdisplaybreaks

\section{Introduction}

In this paper we will review the dark bubble model of a cosmological constant, \cite{Banerjee:2018aa}, and put it into the context of other novel approaches to dark energy. The philosophy behind the program differs in many ways from how the problems of quantum gravity have usually been approached. 

Over the years, supersymmetry has been the guiding principle when looking for structures relevant for model building. This is a fruitful strategy when trying to understand the various ingredients of string theory, and has led to groundbreaking work in mathematics. One has discovered how black holes of various types can have their entropy carried by internal degrees of freedom, suggesting a solution to the black hole information paradox \cite{Strominger:1996sh}. Furthermore, using  compactifications with branes, it has been possible to construct models with low energy physics similar to that of the standard model of particle physics. Further insights have been obtained using the AdS/CFT correspondence and the remarkable, holographic, dualities between gauge theories and gravitational theories. 

Despite all of this progress, dark energy remains elusive. As reviewed in \cite{Danielsson:2018aa}, and the conclusion still stands, there is still not a single trusted example of a metastable de Sitter vacuum in string theory ripe for serious model building. The same is true for inflationary models where it has turned out to be impossible to achieve the required number of e-foldings. The cause of this failure are the notorious vacuum instabilities in any model that cannot rely on supersymmetry.

The swampland program, \cite{Obied:2018sgi}, takes this issue seriously and attempts to find general principles that any string theory realization must satisfy. There has been considerable progress in charting the outskirts of the string landscape, even though its relevance for phenomenology is not clear. Interesting attempts have been made to use this for phenomenology connecting this to the idea of a single extra dimension, the \emph{dark dimension}, which is conjectured to capture the physics of dark energy \cite{Montero:2022prj}. Still, no concrete embedding in string theory has been found. For an overview of the various attempts to construct de Sitter vacua in string theory, see \cite[especially fig. 1]{Berglund:2021ab}, and for a recent, up-to-date review, see \cite{Andriot:2026lac}.

Historically, most approaches have been top down where the theoretical understanding of string theory has been in focus. The hope has been that this will provide us with sufficient understanding to build realistic models of the universe, which would then give testable predictions. The surprising failure to find supersymmetric particles, and the even more surprising discovery of the dark energy in our universe \cite{SupernovaSearchTeam:1998fmf}, show that a different strategy is necessary.

A historical comparison would be the Bohr model of the atom, where a quantitative understanding of the atom was possible even before quantum mechanics was developed. Similarly, a bottom up approach to the cosmological constant problem could yield a stringy model capable of non-trivial quantitative predictions even before a full theoretical underpinning is in place.

Our claim is that the dark bubble is such a model, where we use our current knowledge of what the important structures in string theory are, and put them together to yield a realistic model of the universe. From the simple assumption that it actually works, one can derive a number of non-trivial consistency relations predicting new physics over a wide range of scales. 

The key point is this: the predictions are not mathematical derivations using only first principle string theory. They are derived with the key additional assumption that string theory is a useful theory of the world. In this way we can predict how the theory must behave under conditions as yet out of our mathematical control. To summarize, the question that we try to answer is: {\it If string theory is useful, what properties will the universe have?}  Remarkably, this strategy leads to a number of very concrete results. 

The paper is organized as follows. In \cref{sec:bubbles} we set the stage, identifying the key insights from string theory that we need for our model building. In \cref{sec:dark-dimension} we identify the scales where new phenomena are expected to appear. We will find that the dark bubble is similar to the dark dimension in spirit, though the technical details differ, and in some sense can be viewed as a surprising realization of it. 

In \cref{sec:fat-gravitons} we study 4D gravity in more detail and make predictions for how it needs to change at small (but measurable) distances, and we will discover how gravity shuts itself off at scales smaller than the dark dimension. That is, the dark bubble predicts that gravity becomes {\it weaker} at small length scales.\footnote{This is the opposite of what is expected in a naive realization of the dark dimension.} Amusingly, this is precisely what was argued for in \cite{Sundrum:2003jq} in the context of the fat graviton. The idea is that gravity turns itself off at a scale set by the cosmological constant to remove the usual vacuum energy problem. We will also see how the dark bubble model does not fit into the usual AdS/CFT scenario but instead is a realization of mixed rather than Dirichlet boundary conditions.

In \cref{sec:history-of-universe} we outline the history of the universe making contact with quantum cosmology as well as, in a surprising way, inflation. We conclude with an outlook in \cref{sec:conclusions}.

\section{The Dark Bubble Scenario}\label{sec:bubbles}

The central challenge in cosmology is understanding dark energy. Efforts to describe it within string theory have so far been unsuccessful, mainly because any vacuum state that is not supersymmetric appears to suffer from intrinsic instabilities. Since supersymmetry excludes a positive cosmological constant, our universe cannot be stable in this strict sense.

The swampland program takes the difficulty of constructing a metastable de Sitter vacuum in string theory as a serious clue, and proposes that its absence is tied to profound structural features of the string landscape that are still being investigated. According to this view, any positive vacuum energy must evolve rather than remain constant, and this running is presented as a testable, phenomenological consequence of string theory that awaits experimental verification.

\begin{figure}
    \centering
    \includegraphics[width=0.8\linewidth]{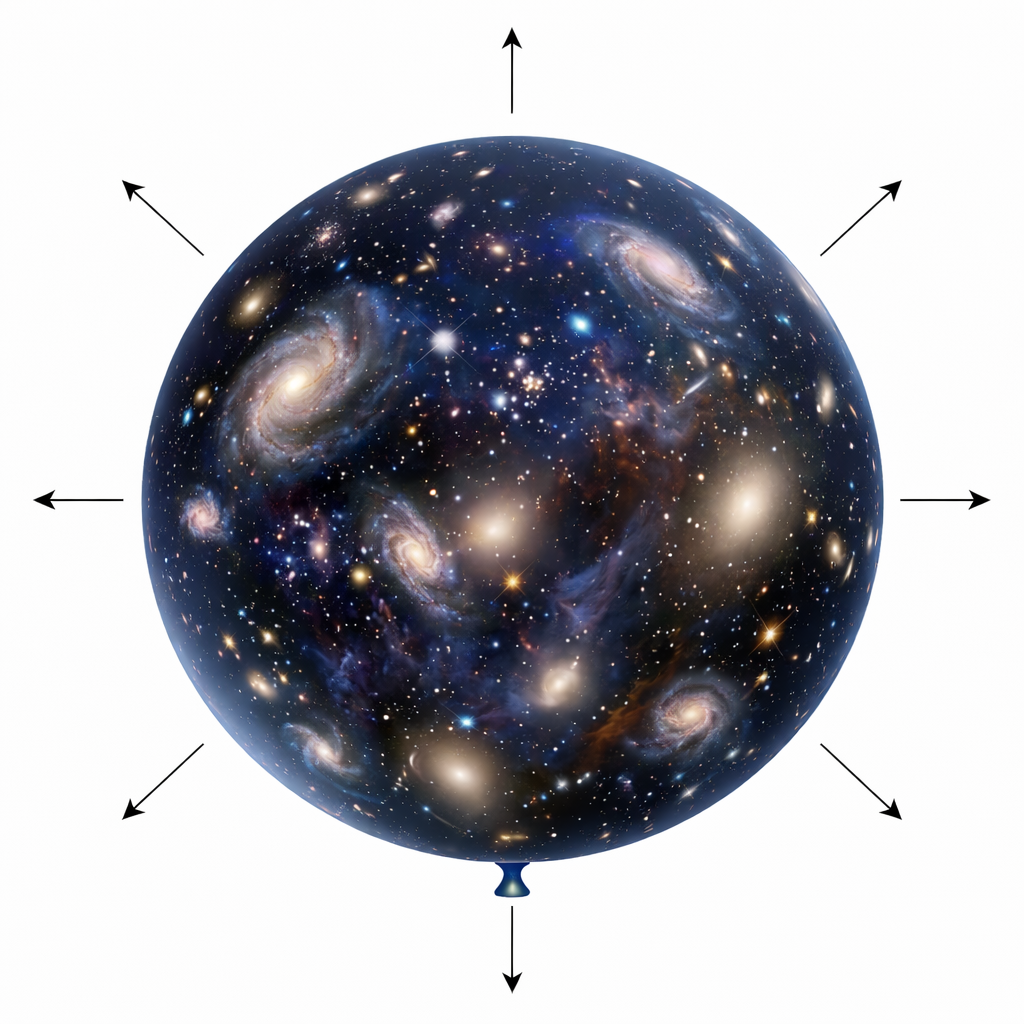}
    \caption{Our universe as an expanding balloon in an extra dimension is literally true in the dark bubble model.}
    \label{fig:balloon}
\end{figure}

The dark bubble model offers an alternative resolution. Rather than forcing string theory into an uncomfortable corner, it embraces these instabilities. Its basic premise is that the familiar popular-science image of our universe as an expanding balloon embedded in an extra dimension is, in fact, literally correct, as depicted in \cref{fig:balloon}.
In usual model building the aim is to establish a stable, and time independent, microscopic 4D vacuum with a positive cosmological constant, which can then be used for cosmology. In the dark bubble model, the microscopic vacuum is an unstable AdS$_5$ vacuum with time dependence built in from the start. The microscopic and macroscopic physics cannot be disentangled, and the stringy instabilities do not prevent the existence of a positive cosmological constant -- they are the cause of it. The different points of view can be thought of as the difference between describing a de Sitter universe using static coordinates, or the time-dependent FRW-coordinates.

Describing our universe as riding a bubble of new vacuum expanding in an extra dimension turns out to be an elegant way to induce a cosmological constant. Concretely, we assume that the transition takes us from one (unstable) AdS$_5$-space to another (more stable). Israel's junction conditions between the two give the energy density on the bubble,
\begin{equation} \label{eq: dark junction}
  \sigma =\frac{3}{8\pi G_5a} \Bigg( \sqrt{k_-^2 a^2+1+ \dot{a}^2}\\
    -\sqrt{k_+^2 a^2+1+ \dot{a}^2}\Bigg) ,
\end{equation}
where $\Lambda_\pm= -6k^2_\pm$ is the negative cosmological constant of the 5D spacetime with $k_-$ on the inside and $k_+$ on the outside, with $k_->k_+$. The radius of the bubble is given by $a(\tau)$, where $\tau$ is proper time on the bubble. In order for the bubble to be able to nucleate, we need the tension of the brane to be slightly below the critical tension, that is $\sigma <\sigma_c \equiv 3(k_--k_+)/(8\pi G_5)$. Assuming $k_\pm a \gg 1$, so that the radius of the universe $a$ is much larger than the AdS-radius $L_\pm = 1/k_\pm $, the second junction condition becomes
\begin{equation}
    \label{eq:Friedmann} 
\frac{\dot{a}^2}{a^2} \approx -\frac{1}{a^2}+\frac{8\pi G_4}{3} \epsilon_\Lambda\,,
\end{equation}
where we have identified the 4D gravitational constant $G_4$, and the cosmological constant $ϵ_Λ$ as
\begin{equation}
\label{eq:G4G5}
    G_4=\frac{2k_-k_+}{k_--k_+}G_5\,, \quad
    \epsilon_\Lambda=\frac{3(k_--k_+)}{8\pi G_5}-\sigma\,.
\end{equation}
We immediately see that a positive cosmological constant is not just a possibility, but inevitable in the induced cosmology on a nucleated bubble. It simply reflects that the effective tension of the brane is less than the critical one, which is necessary for there to be a nucleation in the first place. The key input we need from string theory is the existence of first order phase transitions through bubble nucleation. It is ironic how the dark bubble proposal makes use of what is precisely the problem of the usual approach to finding de Sitter vacua: the notorious instability of any background that is not supersymmetric. Here we see how such an instability in a higher dimension is the direct cause of a positive cosmological constant.

Amusingly, we can already at this point learn something important about how to think of de Sitter space in the context of quantum mechanics. Sometimes it is argued that there is a version of the black hole information paradox also in cosmology. That is, by analogy the cosmological horizon is carrying entropy and there should also exist an associated temperature. It is well known that there is a fundamental difference between the horizon of a black hole and the de Sitter one. The de Sitter horizon is observer dependent, and in many ways similar to the Rindler horizon. While the black hole horizon leads to the information paradox, there is at first glance no such problem associated with the Rindler one. At most it represents an inability to access information, which may be quantified by its area. On the other hand, spherical symmetry makes it difficult to directly map the cosmological horizon to a Rindler one, and it is tempting to associate it with that of a black hole, despite the observer dependence. This problem is solved in the dark bubble model, where the de Sitter temperature can be shown to be simply the Unruh temperature due to the dark bubble accelerating in the fifth dimension \cite{Banerjee:2019aa}. 

Our initial analysis immediately shows why the dark bubble is unique. If the two vacua are close to each other in energy, with $\Delta k /k \ll 1$, where $\Delta k \equiv k_- - k_+$, we find that 4D gravity is much {\it stronger} than 5D gravity. In other words, the 4D Planck length is larger than the 5D Planck length, $l_4>l_5$. In all standard compactifications you find the exact opposite. Using $G_4 \sim G_{4+d}/ L^d$, we find that the lower dimensional Planck length is smaller than the higher dimensional one, that is, $l_4 \sim \left(l_{4+d}/L\right)^{d/2} l_{4+d}<l_{4+d}$ as long as $L>l_{4+d}$. Similarly, a Randall-Sundrum (RS) brane world \cite{Randall:1999aa,Randall:1999ab} has $G^{\textrm{\scshape rs}}_4 =\frac{2k_-k_+}{k_-+k_+}G_5$, which becomes $G^{\textrm{\scshape rs}}_4=kG_5$ when $k_-= k_+ \equiv k$. As long as $L=1/k >l_5$, we again find that the 4D Planck length is smaller than the higher dimensional one. The inversion of this hierarchy is the unique and defining property of the dark bubble model.

Given that the backbone of the dark bubble model is so different from all other string theory constructions, it is not surprising that it does not fit into the pattern of unsuccessful attempts to find de Sitter space in string theory over the past couple of decades that led to the suggestion that there might not exist any de Sitter vacua in string theory \cite{Danielsson:2018aa,Obied:2018sgi}. Therefore, we believe that it is premature to argue that the cosmological constant is part of the swampland. We have, so far, just been looking in the wrong place.

As we have argued, there are two options (except for giving up on string theory). Either, there is no true de Sitter space in string theory and we will soon find that the cosmological constant is changing its value, or we simply have not examined all possibilities. The first option has been popular in the context of the Swampland program, while the second is what the dark bubble model is all about.\footnote{There are already some unconfirmed hints that the cosmological constant may be evolving with time \cite{DESI:2024mwx} (However, also see \cite{Efstathiou:2024xcq}). Even if true, this would not be enough to settle the question.}

This is the main observation. In the rest of this article we will develop the model in detail, calculating how the gravitational force between localized objects reduces to ordinary Einstein gravity at distances larger than the dark dimension, as well as to derive and understand the corrections at even smaller distances. As we will see, there are also a number of other concrete predictions.

\section{The Dark Dimension and the Scale Hierarchy}\label{sec:dark-dimension}

The dark bubble model has five length scales that are of fundamental importance and are all related to each other through a natural number $N$, \cite{Danielsson:2023alz}. These are the cosmological scale $R_{\textrm{\scshape h}}$, the AdS-scale $L$, the string scale $l_s$, the 4D Planck scale $l_4$, and the 5D Planck scale $l_5$. Here we have defined $l_d$ through $G_d=8\pi l_d^{d-2}$, and $l_s$ through $l_s^2=\alpha'$. To understand the hierarchy between these scales, we need four relations. Two of these are the standards ones coming from AdS$_5 \times S^5$:
\begin{eqnarray}
L^4=4\pi g_sN\alpha'^{2}\\
    L^3=\frac{2N^2G_5}{\pi} ,
\end{eqnarray}
which guarantee among other things, that a D3-brane has critical tension. In addition, we have the dark bubble relation between the Planck scales as given by \cref{eq:G4G5}. When applying the junction conditions in 5D across the brane, one needs to make sure that the units used are such that $G_5$ remains constant. As shown in \cite{Danielsson:2023aa,Danielsson:2023alz} it then follows that  
\begin{equation}
    \Delta k = -\frac{2}{3} \frac{\Delta N}{N} k,
\end{equation}
resulting in
\begin{equation}
    G_4=\frac{2k^2}{\Delta k}G_5=\frac{3N}{L}G_5.
\end{equation}
We also find a relation involving the cosmological scale through
\begin{equation}
    \rho_\Lambda=\frac{4}{3\pi g_s L^4}\,,
\end{equation}
where $\rho_\Lambda$ is the energy density of the dark energy and $g_s$ is the string coupling. This was obtained in \cite{Danielsson:2023alz} from an expected correction to the D3-brane action of the form $K^4\sim 1/L^4$, where $K$ is the extrinsic curvature of the embedding into AdS-space.\footnote{The precise numerical coefficient is tricky to derive. See \cite{Danielsson:2023alz} for details.} The effective tension of the D3-brane becomes subcritical in line with the weak gravity conjecture for black branes proposed in \cite{Ooguri:2016pdq}. (Also see \cite{Danielsson:2016mtx} for some further reflections.) 

In this way, the phenomenological observation $\rho_{\Lambda}\sim 1/L^4$, which is at the heart of the dark dimension \cite{Montero:2022prj}, is explained by the failure of the dark bubble to have a flat embedding. This is completely dominated by the extrinsic curvature of the dark bubble in the AdS$_5$. In a sense, one can think of the dark bubble model as a concrete realization of the dark dimension. No other stringy embedding is known to exist.

To uniquely fix all scales, given the cosmological constant in our Universe, we still need to determine what the string coupling is. This can be done in the following way. As we will see in the next section, ordinary matter belonging to the standard model is confined to the world brane,\footnote{With neutrinos as a possible exception.} while gravity and the dark sector live in the bulk. A simple way to introduce matter using the bulk was examined in \cite{Banerjee:2018aa,Banerjee:2019aa}, with more details given in \cite{Banerjee:2020aa,Banerjee:2020ab}. If we allow for a fundamental string stretching along the throat of AdS$_5$ ending on the dark bubble it will pull the the bubble upwards. 

\begin{figure}
    \centering
    \includegraphics[width=\linewidth]{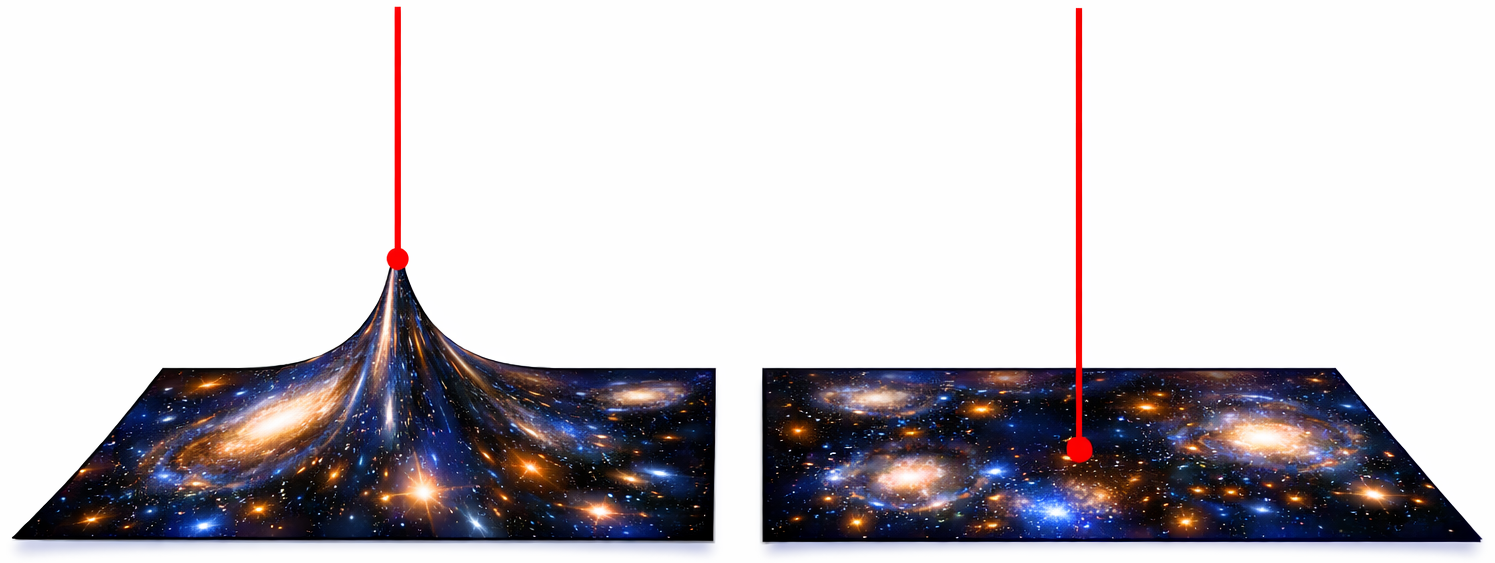}
    \caption{A string pulling on a section of the dark bubble in the bent gauge on the left, and in straight gauge on the right.}
    \label{fig:pulling-string}
\end{figure}

From the point of view of 4D the end point will be interpreted as the presence of a point mass. To figure out its mass we make a coordinate transformation to flat gauge (akin to \cite{Padilla:2004mc}) to figure out what is needed to prevent the string from deforming the brane (see \cref{fig:pulling-string}).

\begin{figure}
    \centering
    \def\svgwidth{\linewidth} %
    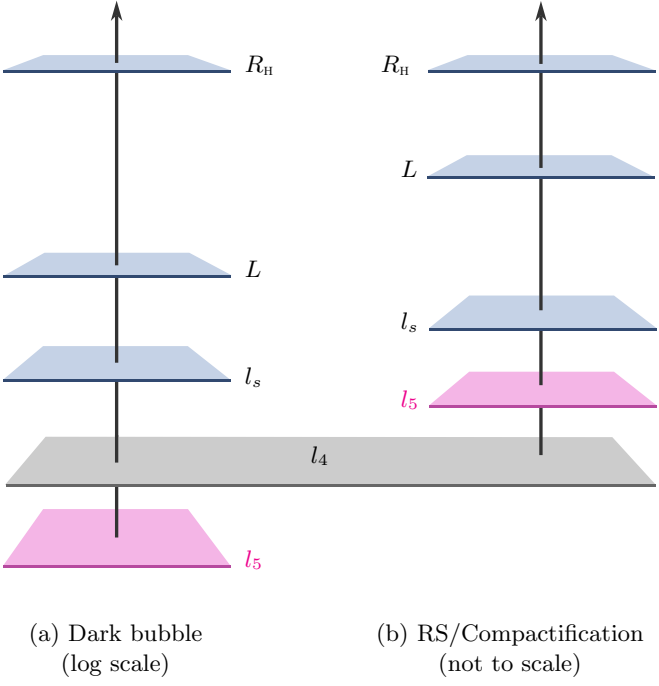 %
    \caption{Hierarchy of fundamental length scales in the dark bubble model compared with RS/standard compactifications. In the dark bubble scenario all scales are fixed by a single parameter $N$ and given by \cref{eq:hierarchy-N}. The scale of the dark dimension, $L$, lies exactly halfway between $l_4$ and $R_{\textrm{\scshape h}}$ in logarithmic units, i.e., $L^2\sim l_4 R_{\textrm{\scshape h}}$, and the string scale lies exactly halfway between $l_4$ and $L$, i.e. $l_s^2 \sim l_4 L$. Notably, in contrast to standard compactifications or RS-braneworlds, for the dark bubble scenario, the 5D Planck length $l_5$ is \emph{smaller} than the 4D Planck length.\label{fig:hierarchy}}
\end{figure}

The required mass becomes \cite{Banerjee:2020ab, Danielsson:2023alz}
\begin{equation}
    M_{p} = T L=\frac{1}{L} \sqrt{\frac{g_sN}{\pi}}=\sqrt{\frac{3 g_s}{2 G_4}},
\end{equation}
that is, of Planckian size.
It is reasonable to interpret the endpoint as a RN-black hole of unit charge, and from this it follows that
\begin{equation}
    M_{{\rm RN}} = \sqrt{\frac{\alpha_{\rm EM}}{G_4}},
\end{equation}
and therefore
\begin{equation}
    g_{s} = \frac{2}{3} \alpha_{\rm EM}.
\end{equation}
With this and the dark energy density in our universe, we obtain $N\sim 10^{63}$, and all scales of the dark bubble model are fixed.\footnote{The pulling string provides an important clue on how to represent black holes in general. If the smallest possible RN-black hole is obtained as a structure that extends out of the brane, why would this not be true for black holes in general? How to describe black holes in the dark bubble model is a subject on its own, see \cite{Banerjee:2021ab,Danielsson:2024frw,Danielsson:2026a}} We now have all the relations that we need, and from this we obtain the following hierarchy of scales as shown in \cref{fig:hierarchy},
\begin{equation}\label{eq:hierarchy}
    l_5 < l_4 < l_s < L < R_{\textrm{\scshape h}}\,,
\end{equation}
with
\begin{equation}\label{eq:hierarchy-N}
     l_5 \sim N^{-1/6} l_4, l_s \sim N^{1/4} l_4, L \sim N^{1/2} l_4, R_{\textrm{\scshape h}} \sim N l_4 .
\end{equation}

From an experimental point of view it is an intriguing result (see \cite{Danielsson:2023alz} for numerical estimates of these energy scales). Not only can we expect modifications of the force of gravity at micron scales, which is the subject of the next section, we also note that the string scale is just above what is currently within reach of the high energy experiments.

\section{Fat Gravitons}\label{sec:fat-gravitons}

Let us now consider matter attached to the brane world represented by the dark bubble, where we expect most of the standard model to live. Just as in general relativity, matter moves around on the world brane according to the equivalence principle in the induced 4D geometry, but the way in which gravity is sourced is more involved. To obtain the effective 4D Einstein equations we combine the junction conditions with a projection of the Gauss-Codazzi equations along the bubble, given by
\begin{align}\nonumber
    \mathcal{J}_{mn} &\equiv R^{(5)}_{αβγδ} e^α_p e^β_m e^γ_q e^δ_n h^{pq}\\
    &= R_{mn} +(K_{mp}K^{p}_n - K K_{mn})\\\nonumber
    &=e_m^\beta e_n^\delta\left(R_{\beta \delta}^{(5)}-R^{(5)}_{\mu \beta \nu \delta} n^\mu n^\nu\right)\,.
\end{align}
From these one can derive the 4D Einstein tensor \cite{Banerjee:2019aa,Banerjee:2022ree}
\begin{align}\nonumber\label{eq:4dEinsteinEq_Exact_small_k}
   G_{ab} &= \frac{k_-k_+}{k_--k_+}\bigg[\left(\frac{\mathcal{J}_{ab}^+}{k_+} - \frac{\mathcal{J}_{ab}^-}{k_-}\right) -\frac{1}{2}\left(\frac{\mathcal{J}^+}{k_+} - \frac{\mathcal{J}^-}{k_-}\right)h_{ab}\\
   &\quad + 3(k_+ - k_-)h_{ab} - 16\pi G_5 S_{ab}\bigg],
\end{align}
where we also have used the second junction condition to replace the extrinsic curvature with the energy momentum tensor of the brane.
Note that this expression is valid in the limit where we only keep terms to linear order in the energy momentum tensor $t_{ab}$, and $S_{ab} \equiv t_{ab} - t h_{ab}$. Corrections will set in at energy densities close to the string scale. We will come back to this in the concluding section on cosmology.

Here we note one of the most important, but also confusing, properties of the dark bubble model. Contrary to RS, the dark bubble has a junction between an inside and an outside. This is much easier to picture than RS, where the brane has two (often identified) insides.\footnote{Equivalently, in Gaussian normal coordinates, with the coordinate $z$ transverse to the bubble wall, the outward pointing normal changes coordinate direction across an ordinary inside/outside interface: schematically $n_\pm = \pm ∂_z$. In an RS-type construction, by contrast, the outward normal points away from the brane into their respective bulk regions: $n_\pm =+∂_z$. This also leads to the important difference in the induced 4D Newton's constant on the dark bubble with the inverted hierarchy.\label{fn:inside-outside}} As we see from \cref{eq:4dEinsteinEq_Exact_small_k}, {\it adding} energy to the brane through $S_{ab}$ {\it reduces} the effective energy momentum tensor in the 4D Einstein equation. This is what is behind the dark bubble understanding of dark energy. If you decrease the tension of the brane, it becomes easier for the brane to nucleate and the positive dark energy becomes larger.\footnote{This was the reason that early works on the dark bubble focused on matter induced from the bulk, such as the stretched strings pulling on the dark bubble \cite{Banerjee:2020aa}.} However, the 4D energy momentum tensor also includes the back reaction from the bulk captured by the first few terms on the right-hand side of \cref{eq:4dEinsteinEq_Exact_small_k}. As we will review, when this is taken into account you obtain a net positive result.

To see how this works, let us solve Einstein's equations in the bulk for a background where the induced metric on the dark bubble is spherically symmetric. We will also make the simplifying assumption that we can ignore the expansion of the universe as well as its positive curvature. Thus we focus on a brane at constant position in Poincare coordinates. Going to Fourier space using
\begin{equation}\label{eq:ht-hp}
    h (r,z) = \frac{1}{r} \sqrt{\frac{2}{\pi}}\int_0^\infty dp\ h(p,z) \sin (p r)\, ,
\end{equation}
the solution of the 5D Einstein equations, to first order in perturbations around AdS, becomes
\begin{equation}\label{eq:hp-solution}
    h (p,z)= A(p) K_2 \left(\frac{p}{k^2 z}\right) + B(p)I_2\left(\frac{p}{k^2 z}\right)\, ,
\end{equation}
 where we focus on the time-time component of the metric, and $A(p), B(p)$ are arbitrary functions. For further details and the full index structure of the metric components, see \cite{Danielsson:2025aa}.
We start with traceless matter and work in a gauge where the brane sits at $z_0$, i.e. it is unbent. 

To satisfy the first junction condition, we match the radial coordinate through $\rho= k_- z_0 r_-=k_+ z_0 r_+$
across the brane,
where $\rho$ is the proper 4D radius on the brane. Similarly, we note that the proper 4D momentum is given by $q=p_-/(k_- z_0)=p_+/(k_+ z_0)$.
We express all momentum dependent functions in terms of $q$ already from the start. The first junction condition then becomes
\begin{equation}
    \label{eq:firstjunction}
    B_+(q) =\frac{A_-(q)K_2(q/k_-)-A_+(q)K_2(q/k_+)}{I_2(q/k_+)}\,,
\end{equation}
where we have used regularity at the center of the bubble to set $B_-(q)=0$.

The second junction condition picks up the presence of energy density on the brane and is given by
\begin{multline}
    \frac{q}{2}\Bigg[B_+(q)I_1\left(\frac{q}{k_+}\right) +A_-(q)K_1 \left(\frac{q}{k_-}\right) \\- A_+(q)K_1 \left(\frac{q}{k_+}\right) \Bigg] = 8 \pi G_5 M(q) ,
\end{multline}
where $M(q)$ is the Fourier sine  transformed energy density.

Since we have three unknown functions $A_\pm(q)$ and $B_+(q)$, we need one more condition to fix all three. This is familiar from ordinary AdS/CFT. Varying the metric at the boundary, that is changing the non-normalizable modes, yields the energy momentum tensor of the CFT as
\begin{equation}
    \delta S_5 \sim \int_{\delta{\cal M}} T^{(CFT)}_{\mu \nu}\delta g^{\mu\nu} \, .
\end{equation}
Choosing $\delta g_{μν}=0$ imposes Dirichlet boundary conditions. In this way, different choices of the non-normalizable modes parametrize different CFT-theories, while the normalizable modes correspond to different expectation values within the same theory. The energy momentum tensor of the CFT is determined by the extrinsic curvature on the cutoff together with counterterms. 

If the non-normalizable terms are dynamical, as they need to be if we want to obtain 4D gravity with a massless graviton, \cite{Banerjee:2023uto}, we can not fix the metric at infinity through Dirichlet boundary conditions. Instead, we need to put a condition on the (renormalized) energy momentum tensor at the boundary. If we were to demand it to vanish, it would correspond to Neumann boundary conditions, but the correct choice for the dark bubble is a mixed boundary condition, where we add a Hilbert term and a cosmological constant term at the cutoff \cite{Danielsson:2025aa}. This is the formal way to pick the relevant boundary conditions, \cite{Compere:2008us,Apostolopoulos:2008ru,Banerjee:2012dw}.

What this amounts to is a balance between the non-normalizable mode against the normalizable one so that 4D gravity remains at the cutoff. The condition is
\begin{equation}\label{eq: mixed}
    B_+(q)=\frac{\eta}{q^2}A_+(q)
\end{equation}
where $\eta$ is a constant determined to be \cite{Danielsson:2025aa}
\begin{equation}
    \eta =4k \Delta k = \frac{8k^3G_5}{G_4}. 
\end{equation}
With this condition, 4D gravity is recovered.

This means that the dark bubble cannot be realized within the boundary conditions associated with AdS/CFT. One can understand this as follows. Let us start with an AdS-space without any nucleated bubble. The spacetime is assumed to be unstable, but with a cutoff not taken all the way to infinity we can, for a sufficiently long lived vacuum, get stability for the time scale that we desire. Within this approximation ordinary AdS/CFT suffices. Eventually a bubble will nucleate, causing a phase transition to a new theory on the boundary, starting at large scales and propagating towards small scales with time. This transition turns on the non-normalizable modes and reflects the presence of a gravitational force induced by the presence of the bubble.

In \cite{Danielsson:2025aa}, the set of conditions derived above were solved in detail, and it was shown that gravity gets weaker at length scales of order $L\sim10^{-5}\rm m$ and effectively shuts itself off. To be precise, we find that the gravitational potential from a point mass is corrected to
\begin{equation}
    V(\rho )= -G_4 M_4 \Bigg[ \frac{1}{\rho}-\frac{3L²}{2\rho^3}+\frac{3L^4\left(31-18\log (ρ/L)\right)}{\rho^5}+...\Bigg].
\end{equation}
The exact expression is somewhat complicated and given in the form of a Fourier transform in \cite{Danielsson:2025aa}.

\begin{figure*}
    \begin{minipage}{0.45\textwidth}
    \centering
    \includegraphics[width=\linewidth]{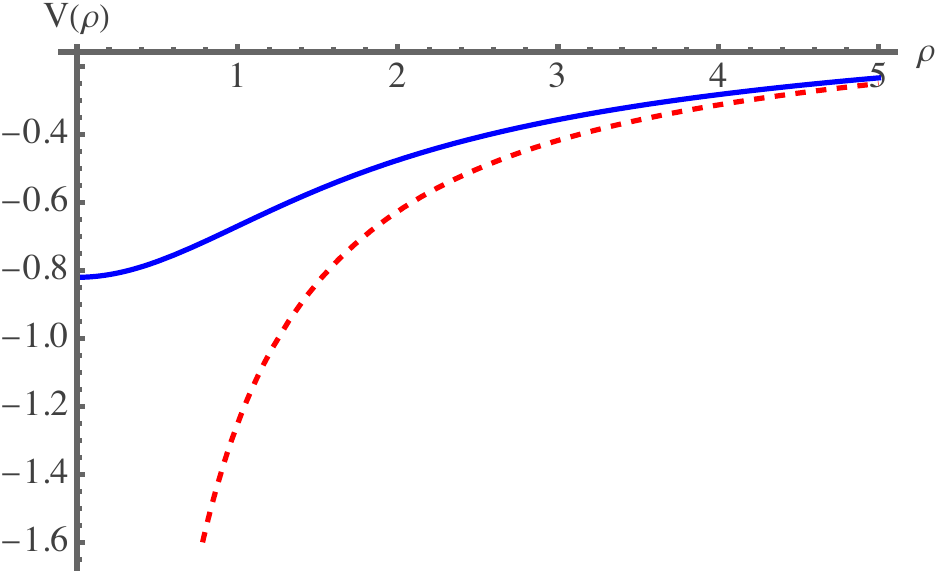}
    {\footnotesize \textbf{(a)} Gravitational potential}
    \end{minipage}
    \hfill
    \begin{minipage}{0.45\textwidth}
    \centering
    \includegraphics[width=\linewidth]{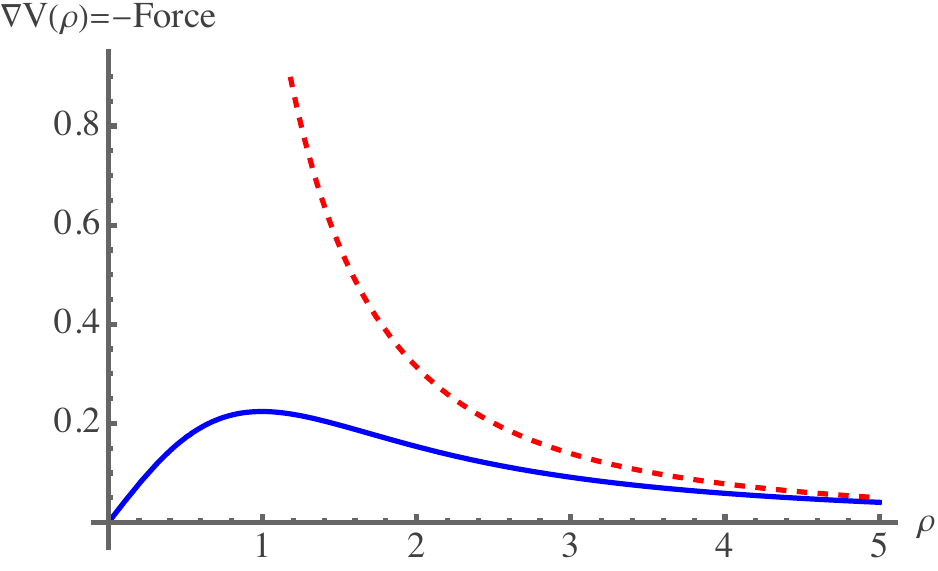}
    {\footnotesize \textbf{(b)} Gravitational force}
    \end{minipage}
    \caption{The gravitational potential and the gravitational force (both in units of $G_4 M_4$) due to a particle of mass $M_4$, as a function of distance from the particle (in units of the AdS$_5$ length $L$). Dashed curves correspond to ordinary 4D gravity, while the solid curves correspond to the 4D universe on a dark bubble. Gravity on the dark bubble gets weaker at small distances. This is exactly what was argued for with a fat graviton in \cite[fig. 8]{Sundrum:2003jq}.\label{fig:pot-force}}
\end{figure*}

The modified gravitational potential and the force are plotted in \cref{fig:pot-force}, where we see how the force is fading away at distances of order $L$. This is exactly the kind of behavior that was suggested in \cite{Sundrum:2003jq}.\footnote{We thank Cumrun Vafa for pointing out this paper to us.} Motivated by the cosmological constant problem, the proposal was that quantum loops of standard model particles would not couple to gravity at energy scales above $\hbar c/L$. That is, gravity cannot probe distances smaller than $L$, which was formulated by saying that gravitons were fat with this size. As a consequence, you would expect $\rho_\Lambda \sim 1/L^4$. In \cite{Sundrum:2003jq} it was argued that all of this could be upheld while still satisfying the equivalence principle for the Standard model matter. 

Another way to argue for gravity becoming weaker at small scales is the non-renormalizability of quantum gravity itself. It is at small scales that these problems show up, and if gravity shuts itself off, the problems would go away. Without any knowledge of high energy physics, you could argue that this is a very natural solution. The question is only, at which scale?  In string theory the usual answer is the string scale far above the scale of the dark dimension. Still, it is ironic that models involving compactification of higher dimensions or Randall-Sundrum like brane-world constructions, always imply a gravitational force that first becomes {\it stronger} in the higher dimensions, before being cutoff near the string scale (see e.g. \cite{Garriga:1999yh}). It is only the dark bubble scenario where gravity gets weaker already at the scale of the dark dimension.

We argue that the \emph{dark bubble is a realization of the fat graviton}, explicitly connecting its presence to the value of the dark energy, and making sure that the non-renormalizability of 4D gravity becomes irrelevant. At small scales gravity fades away and the consistency between quantum mechanics and gravity is assured in the unusual higher dimensional setting of the dark bubble. As explained in \cite{Danielsson:2025aa}, the equivalence principle is guaranteed given that all matter is supposed to follow geodesics in spacetime. That is, the passive gravitational mass and the inertial mass remain the same. Furthermore, Einstein's equations determining the metric are unchanged in the sense that they still relate the Einstein tensor to an effective energy momentum tensor. What differs is that this effective energy momentum tensor sourcing gravity is not identical to the inertial one as measured by a particle physicist. In other words, the active gravitational mass differs from the inertial one in a scale dependent way. That is, they do agree at large scales but at small scales the active energy-momentum tensor is spread out since gravity fails to resolve structures smaller than the dark dimension. This is the fat graviton effect. As shown in \cite{Danielsson:2025aa} all of this can be captured by a scale dependent Newton's constant.\footnote{As further explained in \cite{Danielsson:2025aa} its universal form, independent of the type of matter, guarantees that the Newton's third law holds.}

The dark bubble model provides further insight into the role the dark dimension is playing. It is not only a direction in which the universe is expanding and exploring, it also a holographic representation of the energy scales. This is tightly connected to the mixed boundary condition that we had to impose. The non-renormalizability of gravity does not imply that the theory is inconsistent, just that it lacks predictability due to all counterterms that need to be added to cancel all the divergences. It was suggested in \cite{Danielsson:2025aa}, that this is captured by the need of the extra boundary condition at holographic infinity in the dark bubble model. What saves the day is the presence of a natural mixed boundary condition, \cref{eq: mixed}, which uniquely specifies the theory. It basically amounts to the assumption that there are no new surprising structures in 5D space that the expanding bubble will encounter in the future, or, equivalently, the high energy behavior of the dark sector is a desert.

\section{The History of the Universe}\label{sec:history-of-universe}

We will now shift focus back to the first Friedmann equation of a FRW-cosmology as given by the second junction condition for the dark bubble. In \cref{eq: dark junction}, we made an expansion assuming that the Hubble constant was small enough for higher order corrections to be ignored. What if we keep all terms? As we will see, there will be important corrections at high energy densities with dramatic consequences, \cite{Danielsson:2025aa}.

The equation of motion that governs the expansion of the universe, obtained from the full junction conditions, is given by 
\begin{multline}
    \sigma + \rho_r=\frac{3}{8\pi G_5a} \Bigg( \sqrt{k_-^2 a^2+1+ \dot{a}^2}\\
    -\sqrt{k_+^2 a^2+1+ \dot{a}^2-\frac{2G_5 M_+}{a^2}}\Bigg) .
\end{multline}
Here we have included radiation on the brane, $\rho_r \sim 1/a^4$, as well as a bulk backreaction on the outside in the form of an AdS-Schwarzschild metric.
For small $\Delta k$, it can be brought to the form of a Friedmann equation given by
\begin{equation} \label{eq:corrFriedman}
    \frac{\dot{a}^2}{a^2}=-\frac{1}{a^2}+k^2\left( \frac{1+\frac{G_4M_+}{2k^3 a^4}}{1 +\frac{G_4M_-}{ 2k^3 a^4}}+\frac{G_5M_-}{2k^2 a^4}\right)^2-k^2
\end{equation}
At late times, when $a$ is large, we find 
\begin{equation}\label{eq:Friedmann-eq0}
\frac{\dot{a}^2}{a^2}=-\frac{1}{a^2}+\frac{G_4(M_+-M_-)}{ka^4}=-\frac{1}{a^2}+\frac{8\pi G_4}{3}\rho_r \, ,
\end{equation}
provided that we make the choice $M_+=2M_-$, which is equivalent to the choices made earlier for localized matter. Note that for a homogeneous matter distribution, the mixed boundary condition, with $q\rightarrow 0$, tells us that there will be only normalizable modes on the outside. This corresponds to the AdS-Schwarzschild metric.

At small $a$, the Friedmann equation becomes $H² = 3 k²$,
that is, the effective energy density seen by gravity is constant in time, even though the inertial energy density relevant to a particle physicist keeps growing towards smaller values of $a$. This implies that the early universe undergoes a phase of inflation with the Hubble constant given by $H \sim 1/L$ that lasts around 30 e-folds \cite{Danielsson:2025aa}. The surprising behavior is a reflection of the weakening of gravity that we saw earlier. Interestingly, a similar phenomenon was argued for in \cite{Khoury:2006fg} inspired by the fat graviton.

As we go even further back in time we reach Planckian energy densities coming to the very origin of our universe. To describe the actual nucleation we need to write the junction conditions introducing a canonical momentum so that we can perform a canonical quantization. We find
\begin{multline}
    H=2\pi^3 r^2 \sigma-\frac{3\pi^2 r^2}{4G_5}\Bigg[\left(f_-(r)^{1/2}-f_+(r)^{1/2}\right)^2\\
    -2\sqrt{f_-(r)f_+(r)}\left( 1-\cosh \frac{4G_5p}{3\pi^2r^2}\right)\Bigg]^{1/2}
\end{multline}
with $H=0$, where $f_\pm(r)=k_\pm^2r^2+1$. In \cite{Danielsson:2021aa} it was shown, for a universe without radiation, how the Hamiltonian in the limit of small $p$ becomes quadratic in the momentum, and how the Schrödinger equation $\hat{H}\psi(r)=0$ becomes identical to the Wheeler-de Witt equation of Vilenkin's quantum cosmology. In the dark bubble model the tunneling event is interpreted as the nucleation of an expanding bubble in 5D.

For general values of $p$ the Hamiltonian constraint turns into a difference equation. This might seem surprising, but has very important consequences. Such equations are common in condensed matter physics, where they describe electrons moving in the lattice structure of crystals, leading to a band structure. Here, we see that the discreteness corresponds to a quantization of the volume of 4D in units of $G_5$. This differs from the common guess that quantization should be in units of size $l_4^3 \sim G_4^{3/2}$. Since spacetime is a brane embedded into 5D, it is not surprising that it inherits the discretization associated with the horizon of a 5D black hole.~\footnote{A bubble nucleating in 4D has its area quantized in units of $G_4$ by the same reasoning. This plays a role in interpreting black shells as black hole mimickers, \cite{Danielsson:2026b}. It is remarkable that this simple analysis of quantum bubble nucleation in different dimensions leads to a discretization of space at the Planck scale. Area quantization of black holes has previously been suggested in \cite{Bekenstein:1974jk,Mukhanov:1986me,Bekenstein:1995ju} based on heuristic considerations. It is intriguing that the dark bubble and black shell framework may be pointing toward a similar type of quantization.}

The nucleation of a universe with matter proceeds in a different way. Contrary to the standard inflationary scenario, the matter in the dark bubble universe is present from the very beginning.\footnote{Even if there is an inflationary phase, this is not driven by dark energy that gives rise to reheating.} In \cite{Danielsson:2025aa}, we proposed that it all starts with a black hole of a critical size in AdS$_5$. Such a black hole might be the result of a collapsing cloud of matter in AdS$_5$. As the black hole grows in size while accreting matter, a critical point is reached when the horizon radius is of order $L$, and it becomes possible for the dark bubble to nucleate at its horizon. If the black hole is just a little larger, the barrier disappears and the black hole is classically unstable against the nucleation. The matter of the original black hole becomes matter on top of the dark bubble. This is the creation of our universe.

At this moment, the energy density of the universe is $\frac{L^2}{G_5}/L^3 \sim 1/l_4^4$, i.e., of order the 4D Planck scale.~\footnote{Here and throughout, we make use of the scalings in \cref{eq:hierarchy-N}} As mentioned earlier, however, spacetime is not discretized at the 4D Planck length $l_4$, but rather at the much smaller 5D Planck length $l_5 \ll l_4$. Consequently, the entropy content of the universe is $L^3/l_5^3 \sim N^2$. This energy density is consistent with a configuration consisting of (in addition to the single dark bubble brane) a mixture of $N$ brane–antibrane pairs, whose total energy density is $\sim N/l_s^4 \sim 1/l_4^4$, together with $N^2$ massless modes at temperature $1/L$, giving an energy density $N^2/L^4 \sim 1/l_4^4$, of the same order as that stored in the branes.

Immediately following nucleation, the universe undergoes an exponential expansion characterized by a Hubble parameter of order $1/L$. Because gravity is relatively weak, with $\tilde{G}_4 = G_4/N$, the cosmological horizon, whose area is $L^2$, can still accommodate all the required degrees of freedom that have crossed it, yielding $L^2/\tilde{G}_4 \sim N^2$. This phase persists until the energy density falls from the Planck scale, $\sim 1/l_4^4$, down to the string scale, $\sim 1/l_s^4$, during which the universe grows by a factor of $N^{1/4}$. Once the energy density reaches the string scale, no brane–antibrane pairs remain, and all the energy must be carried by the matter associated with the single surviving standard model dark bubble brane.

By this stage, the strength of 4D gravity has grown from $kG_5$ to $G_4 \sim N k G_5$. This reduces the entropy of the horizon to $L^2/G_4 \sim N \ll N^2$. This does not signal any inconsistency. All the degrees of freedom, and thus the associated entropy, remain present at large (infrared) scales. During the radiation-dominated era, as the horizon expands, an increasing number of modes that exited during inflation re-enter. Eventually, at relatively recent times (redshift of order a few), the universe once more transitioned into an accelerated expansion phase. Degrees of freedom then begin to exit again, and the final entropy accounting is performed using the cosmological horizon, which once more yields $R_{\textrm{\scshape h}}^2/G_4 \sim N^2$. 

We now assume that most of the energy stored in the original $N^2$ degrees of freedom is converted into ordinary radiation that is part of the Standard Model. A small fraction of this radiation is later converted into matter, while nearly all of the radiation observed today, with a magnitude of $\Omega_{\rm rad} \sim 5 \times 10^{-5}$, originates from these stringy energy densities.

We now arrive at the model’s most striking prediction. Since essentially all of the radiation in today’s universe can be traced back to its initial nucleation, the total mass content should be determined by the mass of the original 5D black hole. Although its radius is microscopic, of order $L$, the key point specific to the dark bubble is that 5D gravity is extremely weak, so the corresponding mass is in fact enormous. One obtains $M \sim L^2/G_5 \sim N^{3/2}/l_4$. This mass initially appears as radiation, and by the time the universe has expanded to a radius $R_{\textrm{\scshape h}}$, i.e. by a factor $N^{1/2}$, the total energy has been redshifted down to $N/l_4 \sim R_{\textrm{\scshape h}}^3 \times 1/L^4$. At that epoch, the radiation energy density is therefore of the same order as the dark energy density. This no longer holds today, implying that the current universe must be larger by a few orders of magnitude. Numerically, it was found in \cite{Danielsson:2025aa} that
\begin{equation}
    \Omega_{\rm curv} \sim 5 \times 10^{-4}\,,
\end{equation}
which is, quite remarkably, just below the present observational limits.

\section{Conclusions}\label{sec:conclusions}

Let us summarize the key features of the dark-bubble model and its main phenomenological implications.

The fundamental assumption is that our universe is riding an expanding spherical brane bubble in 5D AdS, formed through a first-order phase transition. The observed accelerated expansion is a consequence of the bubble tension being below a critical threshold; in four-dimensional terms, this manifests as an effective 4D cosmological constant. As a brane world, the dark bubble differs fundamentally from the Randall–Sundrum (RS) scenario, in which two AdS regions are glued together such that their common boundary has two “insides”. In contrast, the expanding dark bubble has a genuine inside and an outside (cf. footnote~\ref{fn:inside-outside}). 

Applying the junction conditions together with the Gauss–Codazzi equations, determines the effective 4D Newton’s constant. A central and distinctive result is that four dimensional gravity is stronger than the five dimensional one, inverting the usual hierarchy found in standard Kaluza-Klein compactifications or RS-type setups. For 4D gravity to emerge, the non-normalizable modes must be dynamical, which requires mixed boundary conditions in AdS. This stands in contrast to the AdS/CFT framework, where boundary conditions are taken to be Dirichlet.

The construction leads to several observational consequences, of which we highlight a few. First, a positive cosmological constant is a defining property of the model. Its size is directly tied to the existence of a dark extra dimension with an AdS length scale of order microns. At distances shorter than this scale, 4D gravitational interactions are strongly suppressed and effectively shut off. The same mechanism simultaneously realizes the dark dimension scenario of \cite{Montero:2022prj}, and the fat graviton proposal of \cite{Sundrum:2003jq}. In this sense, the dark bubble provides a concrete realization of both ideas within a single dynamical framework, where the weakening of gravity at short distances is directly tied to the cosmological constant scale. The model also predicts a surprisingly low energy for the string scale, which is expected to lie around tens of TeV, pointing to new physics just beyond the current reach of current experiments

That said, the dark bubble framework also raises a number of important open questions. Most significantly, a fully controlled string theory embedding is still lacking. Such an embedding will provide a deeper understanding of the assumptions underlying the model and its connection to the broader string theory landscape, taking the dark bubble from being a “scenario” to a “model”. 
Equally important is the realization of the Standard model of particle physics within this framework. While matter localized on the bubble can be incorporated, a derivation of the full particle spectrum, gauge structure and interactions from the underlying higher dimensional theory remains to be achieved. Some steps in this direction have been taken in \cite{Basile:2023aa,Danielsson:2026qch} in the case of electromagnetic waves. The neutrino sector could be of special interest, since its mass scale is close to the dark dimension scale. This suggests that neutrinos might serve as a bridge between matter on the brane and in the bulk.

Finally, the role of mixed boundary conditions deserves further clarification. Although they are crucial for obtaining 4D gravity and arise naturally in the present construction, it would be desirable to derive them from first principles in string theory rather than imposing them at the level of the effective description. In particular, a deeper understanding of how these boundary conditions emerge in a time-dependent setting may be essential for clarifying the holographic interpretation of the dark bubble and its relation to more conventional AdS/CFT dualities.

In summary, \emph{the dark bubble model provides a unified framework in which a positive cosmological constant, a micron-scale extra “dark” dimension, and a suppression of gravity at short distances arise naturally}. At the same time, it highlights several directions where a more fundamental microscopic realization is required--most notably in string embedding, Standard model of particle physics realization, and the origin of the mixed boundary conditions.

\section*{Acknowledgements}
 We thank Alessandro Tomasiello, Cumrun Vafa, Vincent Van Hemelryck, and Thomas Van Riet for discussions, and Kungliga Fysiografiska sällskapet i Lund for support.

\bibliography{refs}

\begin{thebibliography}{39}%
\makeatletter
\providecommand \@ifxundefined [1]{%
 \@ifx{#1\undefined}
}%
\providecommand \@ifnum [1]{%
 \ifnum #1\expandafter \@firstoftwo
 \else \expandafter \@secondoftwo
 \fi
}%
\providecommand \@ifx [1]{%
 \ifx #1\expandafter \@firstoftwo
 \else \expandafter \@secondoftwo
 \fi
}%
\providecommand \natexlab [1]{#1}%
\providecommand \enquote  [1]{``#1''}%
\providecommand \bibnamefont  [1]{#1}%
\providecommand \bibfnamefont [1]{#1}%
\providecommand \citenamefont [1]{#1}%
\providecommand \href@noop [0]{\@secondoftwo}%
\providecommand \href [0]{\begingroup \@sanitize@url \@href}%
\providecommand \@href[1]{\@@startlink{#1}\@@href}%
\providecommand \@@href[1]{\endgroup#1\@@endlink}%
\providecommand \@sanitize@url [0]{\catcode `\\12\catcode `\$12\catcode
  `\&12\catcode `\#12\catcode `\^12\catcode `\_12\catcode `\%12\relax}%
\providecommand \@@startlink[1]{}%
\providecommand \@@endlink[0]{}%
\providecommand \url  [0]{\begingroup\@sanitize@url \@url }%
\providecommand \@url [1]{\endgroup\@href {#1}{\urlprefix }}%
\providecommand \urlprefix  [0]{URL }%
\providecommand \Eprint [0]{\href }%
\providecommand \doibase [0]{https://doi.org/}%
\providecommand \selectlanguage [0]{\@gobble}%
\providecommand \bibinfo  [0]{\@secondoftwo}%
\providecommand \bibfield  [0]{\@secondoftwo}%
\providecommand \translation [1]{[#1]}%
\providecommand \BibitemOpen [0]{}%
\providecommand \bibitemStop [0]{}%
\providecommand \bibitemNoStop [0]{.\EOS\space}%
\providecommand \EOS [0]{\spacefactor3000\relax}%
\providecommand \BibitemShut  [1]{\csname bibitem#1\endcsname}%
\let\auto@bib@innerbib\@empty
\bibitem [{\citenamefont {Banerjee}\ \emph {et~al.}(2018)\citenamefont
  {Banerjee}, \citenamefont {Danielsson}, \citenamefont {Dibitetto},
  \citenamefont {Giri},\ and\ \citenamefont {Schillo}}]{Banerjee:2018aa}%
  \BibitemOpen
  \bibfield  {author} {\bibinfo {author} {\bibfnamefont {S.}~\bibnamefont
  {Banerjee}}, \bibinfo {author} {\bibfnamefont {U.}~\bibnamefont
  {Danielsson}}, \bibinfo {author} {\bibfnamefont {G.}~\bibnamefont
  {Dibitetto}}, \bibinfo {author} {\bibfnamefont {S.}~\bibnamefont {Giri}},\
  and\ \bibinfo {author} {\bibfnamefont {M.}~\bibnamefont {Schillo}},\
  }\bibfield  {title} {\bibinfo {title} {{Emergent de Sitter Cosmology from
  Decaying Anti\textendash{}de Sitter Space}},\ }\href
  {https://doi.org/10.1103/PhysRevLett.121.261301} {\bibfield  {journal}
  {\bibinfo  {journal} {Phys. Rev. Lett.}\ }\textbf {\bibinfo {volume} {121}},\
  \bibinfo {pages} {261301} (\bibinfo {year} {2018})},\ \Eprint
  {https://arxiv.org/abs/1807.01570} {arXiv:1807.01570 [hep-th]} \BibitemShut
  {NoStop}%
\bibitem [{\citenamefont {Strominger}\ and\ \citenamefont
  {Vafa}(1996)}]{Strominger:1996sh}%
  \BibitemOpen
  \bibfield  {author} {\bibinfo {author} {\bibfnamefont {A.}~\bibnamefont
  {Strominger}}\ and\ \bibinfo {author} {\bibfnamefont {C.}~\bibnamefont
  {Vafa}},\ }\bibfield  {title} {\bibinfo {title} {{Microscopic origin of the
  Bekenstein-Hawking entropy}},\ }\href
  {https://doi.org/10.1016/0370-2693(96)00345-0} {\bibfield  {journal}
  {\bibinfo  {journal} {Phys. Lett. B}\ }\textbf {\bibinfo {volume} {379}},\
  \bibinfo {pages} {99} (\bibinfo {year} {1996})},\ \Eprint
  {https://arxiv.org/abs/hep-th/9601029} {arXiv:hep-th/9601029} \BibitemShut
  {NoStop}%
\bibitem [{\citenamefont {Danielsson}\ and\ \citenamefont
  {Van~Riet}(2018)}]{Danielsson:2018aa}%
  \BibitemOpen
  \bibfield  {author} {\bibinfo {author} {\bibfnamefont {U.~H.}\ \bibnamefont
  {Danielsson}}\ and\ \bibinfo {author} {\bibfnamefont {T.}~\bibnamefont
  {Van~Riet}},\ }\bibfield  {title} {\bibinfo {title} {{What if string theory
  has no de Sitter vacua?}},\ }\href
  {https://doi.org/10.1142/S0218271818300070} {\bibfield  {journal} {\bibinfo
  {journal} {Int. J. Mod. Phys. D}\ }\textbf {\bibinfo {volume} {27}},\
  \bibinfo {pages} {1830007} (\bibinfo {year} {2018})},\ \Eprint
  {https://arxiv.org/abs/1804.01120} {arXiv:1804.01120 [hep-th]} \BibitemShut
  {NoStop}%
\bibitem [{\citenamefont {Obied}\ \emph {et~al.}(2018)\citenamefont {Obied},
  \citenamefont {Ooguri}, \citenamefont {Spodyneiko},\ and\ \citenamefont
  {Vafa}}]{Obied:2018sgi}%
  \BibitemOpen
  \bibfield  {author} {\bibinfo {author} {\bibfnamefont {G.}~\bibnamefont
  {Obied}}, \bibinfo {author} {\bibfnamefont {H.}~\bibnamefont {Ooguri}},
  \bibinfo {author} {\bibfnamefont {L.}~\bibnamefont {Spodyneiko}},\ and\
  \bibinfo {author} {\bibfnamefont {C.}~\bibnamefont {Vafa}},\ }\bibfield
  {title} {\bibinfo {title} {{De Sitter Space and the Swampland}},\ }\href@noop
  {} {\  (\bibinfo {year} {2018})},\ \Eprint {https://arxiv.org/abs/1806.08362}
  {arXiv:1806.08362 [hep-th]} \BibitemShut {NoStop}%
\bibitem [{\citenamefont {Montero}\ \emph {et~al.}(2023)\citenamefont
  {Montero}, \citenamefont {Vafa},\ and\ \citenamefont
  {Valenzuela}}]{Montero:2022prj}%
  \BibitemOpen
  \bibfield  {author} {\bibinfo {author} {\bibfnamefont {M.}~\bibnamefont
  {Montero}}, \bibinfo {author} {\bibfnamefont {C.}~\bibnamefont {Vafa}},\ and\
  \bibinfo {author} {\bibfnamefont {I.}~\bibnamefont {Valenzuela}},\ }\bibfield
   {title} {\bibinfo {title} {{The dark dimension and the Swampland}},\ }\href
  {https://doi.org/10.1007/JHEP02(2023)022} {\bibfield  {journal} {\bibinfo
  {journal} {JHEP}\ }\textbf {\bibinfo {volume} {02}},\ \bibinfo {pages}
  {022}},\ \Eprint {https://arxiv.org/abs/2205.12293} {arXiv:2205.12293
  [hep-th]} \BibitemShut {NoStop}%
\bibitem [{\citenamefont {Berglund}\ \emph {et~al.}(2021)\citenamefont
  {Berglund}, \citenamefont {H\"ubsch},\ and\ \citenamefont
  {Minic}}]{Berglund:2021ab}%
  \BibitemOpen
  \bibfield  {author} {\bibinfo {author} {\bibfnamefont {P.}~\bibnamefont
  {Berglund}}, \bibinfo {author} {\bibfnamefont {T.}~\bibnamefont {H\"ubsch}},\
  and\ \bibinfo {author} {\bibfnamefont {D.}~\bibnamefont {Minic}},\ }\bibfield
   {title} {\bibinfo {title} {{Stringy Bubbles Solve de Sitter Troubles}},\
  }\href {https://doi.org/10.3390/universe7100363} {\bibfield  {journal}
  {\bibinfo  {journal} {Universe}\ }\textbf {\bibinfo {volume} {7}},\ \bibinfo
  {pages} {363} (\bibinfo {year} {2021})},\ \Eprint
  {https://arxiv.org/abs/2109.01122} {arXiv:2109.01122 [hep-th]} \BibitemShut
  {NoStop}%
\bibitem [{\citenamefont {Andriot}(2026)}]{Andriot:2026lac}%
  \BibitemOpen
  \bibfield  {author} {\bibinfo {author} {\bibfnamefont {D.}~\bibnamefont
  {Andriot}},\ }\bibfield  {title} {\bibinfo {title} {{Dark energy from string
  theory: an introductory review}},\ }\href@noop {} {\  (\bibinfo {year}
  {2026})},\ \Eprint {https://arxiv.org/abs/2603.25797} {arXiv:2603.25797
  [hep-th]} \BibitemShut {NoStop}%
\bibitem [{\citenamefont {Riess}\ \emph {et~al.}(1998)\citenamefont {Riess}
  \emph {et~al.}}]{SupernovaSearchTeam:1998fmf}%
  \BibitemOpen
  \bibfield  {author} {\bibinfo {author} {\bibfnamefont {A.~G.}\ \bibnamefont
  {Riess}} \emph {et~al.} (\bibinfo {collaboration} {Supernova Search Team}),\
  }\bibfield  {title} {\bibinfo {title} {{Observational evidence from
  supernovae for an accelerating universe and a cosmological constant}},\
  }\href {https://doi.org/10.1086/300499} {\bibfield  {journal} {\bibinfo
  {journal} {Astron. J.}\ }\textbf {\bibinfo {volume} {116}},\ \bibinfo {pages}
  {1009} (\bibinfo {year} {1998})},\ \Eprint
  {https://arxiv.org/abs/astro-ph/9805201} {arXiv:astro-ph/9805201}
  \BibitemShut {NoStop}%
\bibitem [{\citenamefont {Sundrum}(2004)}]{Sundrum:2003jq}%
  \BibitemOpen
  \bibfield  {author} {\bibinfo {author} {\bibfnamefont {R.}~\bibnamefont
  {Sundrum}},\ }\bibfield  {title} {\bibinfo {title} {{Fat gravitons, the
  cosmological constant and submillimeter tests}},\ }\href
  {https://doi.org/10.1103/PhysRevD.69.044014} {\bibfield  {journal} {\bibinfo
  {journal} {Phys. Rev. D}\ }\textbf {\bibinfo {volume} {69}},\ \bibinfo
  {pages} {044014} (\bibinfo {year} {2004})},\ \Eprint
  {https://arxiv.org/abs/hep-th/0306106} {arXiv:hep-th/0306106} \BibitemShut
  {NoStop}%
\bibitem [{\citenamefont {Banerjee}\ \emph {et~al.}(2019)\citenamefont
  {Banerjee}, \citenamefont {Danielsson}, \citenamefont {Dibitetto},
  \citenamefont {Giri},\ and\ \citenamefont {Schillo}}]{Banerjee:2019aa}%
  \BibitemOpen
  \bibfield  {author} {\bibinfo {author} {\bibfnamefont {S.}~\bibnamefont
  {Banerjee}}, \bibinfo {author} {\bibfnamefont {U.}~\bibnamefont
  {Danielsson}}, \bibinfo {author} {\bibfnamefont {G.}~\bibnamefont
  {Dibitetto}}, \bibinfo {author} {\bibfnamefont {S.}~\bibnamefont {Giri}},\
  and\ \bibinfo {author} {\bibfnamefont {M.}~\bibnamefont {Schillo}},\
  }\bibfield  {title} {\bibinfo {title} {{de Sitter Cosmology on an expanding
  bubble}},\ }\href {https://doi.org/10.1007/JHEP10(2019)164} {\bibfield
  {journal} {\bibinfo  {journal} {JHEP}\ }\textbf {\bibinfo {volume} {10}},\
  \bibinfo {pages} {164}},\ \Eprint {https://arxiv.org/abs/1907.04268}
  {arXiv:1907.04268 [hep-th]} \BibitemShut {NoStop}%
\bibitem [{\citenamefont {Randall}\ and\ \citenamefont
  {Sundrum}(1999{\natexlab{a}})}]{Randall:1999aa}%
  \BibitemOpen
  \bibfield  {author} {\bibinfo {author} {\bibfnamefont {L.}~\bibnamefont
  {Randall}}\ and\ \bibinfo {author} {\bibfnamefont {R.}~\bibnamefont
  {Sundrum}},\ }\bibfield  {title} {\bibinfo {title} {Large mass hierarchy from
  a small extra dimension},\ }\href
  {https://doi.org/10.1103/physrevlett.83.3370} {\bibfield  {journal} {\bibinfo
   {journal} {Physical Review Letters}\ }\textbf {\bibinfo {volume} {83}},\
  \bibinfo {pages} {3370–3373} (\bibinfo {year} {1999}{\natexlab{a}})},\
  \Eprint {https://arxiv.org/abs/9905221} {arXiv:9905221 [hep-ph]} \BibitemShut
  {NoStop}%
\bibitem [{\citenamefont {Randall}\ and\ \citenamefont
  {Sundrum}(1999{\natexlab{b}})}]{Randall:1999ab}%
  \BibitemOpen
  \bibfield  {author} {\bibinfo {author} {\bibfnamefont {L.}~\bibnamefont
  {Randall}}\ and\ \bibinfo {author} {\bibfnamefont {R.}~\bibnamefont
  {Sundrum}},\ }\bibfield  {title} {\bibinfo {title} {An alternative to
  compactification},\ }\href {https://doi.org/10.1103/physrevlett.83.4690}
  {\bibfield  {journal} {\bibinfo  {journal} {Physical Review Letters}\
  }\textbf {\bibinfo {volume} {83}},\ \bibinfo {pages} {4690–4693} (\bibinfo
  {year} {1999}{\natexlab{b}})},\ \Eprint {https://arxiv.org/abs/9906064}
  {arXiv:9906064 [hep-th]} \BibitemShut {NoStop}%
\bibitem [{\citenamefont {Adame}\ \emph {et~al.}(2025)\citenamefont {Adame}
  \emph {et~al.}}]{DESI:2024mwx}%
  \BibitemOpen
  \bibfield  {author} {\bibinfo {author} {\bibfnamefont {A.~G.}\ \bibnamefont
  {Adame}} \emph {et~al.} (\bibinfo {collaboration} {DESI}),\ }\bibfield
  {title} {\bibinfo {title} {{DESI 2024 VI: cosmological constraints from the
  measurements of baryon acoustic oscillations}},\ }\href
  {https://doi.org/10.1088/1475-7516/2025/02/021} {\bibfield  {journal}
  {\bibinfo  {journal} {JCAP}\ }\textbf {\bibinfo {volume} {02}},\ \bibinfo
  {pages} {021}},\ \Eprint {https://arxiv.org/abs/2404.03002} {arXiv:2404.03002
  [astro-ph.CO]} \BibitemShut {NoStop}%
\bibitem [{\citenamefont {Efstathiou}(2025)}]{Efstathiou:2024xcq}%
  \BibitemOpen
  \bibfield  {author} {\bibinfo {author} {\bibfnamefont {G.}~\bibnamefont
  {Efstathiou}},\ }\bibfield  {title} {\bibinfo {title} {{Evolving dark energy
  or supernovae systematics?}},\ }\href {https://doi.org/10.1093/mnras/staf301}
  {\bibfield  {journal} {\bibinfo  {journal} {Mon. Not. Roy. Astron. Soc.}\
  }\textbf {\bibinfo {volume} {538}},\ \bibinfo {pages} {875} (\bibinfo {year}
  {2025})},\ \Eprint {https://arxiv.org/abs/2408.07175} {arXiv:2408.07175
  [astro-ph.CO]} \BibitemShut {NoStop}%
\bibitem [{\citenamefont {Danielsson}\ and\ \citenamefont
  {Panizo}(2024)}]{Danielsson:2023alz}%
  \BibitemOpen
  \bibfield  {author} {\bibinfo {author} {\bibfnamefont {U.}~\bibnamefont
  {Danielsson}}\ and\ \bibinfo {author} {\bibfnamefont {D.}~\bibnamefont
  {Panizo}},\ }\bibfield  {title} {\bibinfo {title} {{Experimental tests of
  dark bubble cosmology}},\ }\href
  {https://doi.org/10.1103/PhysRevD.109.026003} {\bibfield  {journal} {\bibinfo
   {journal} {Phys. Rev. D}\ }\textbf {\bibinfo {volume} {109}},\ \bibinfo
  {pages} {026003} (\bibinfo {year} {2024})},\ \Eprint
  {https://arxiv.org/abs/2311.14589} {arXiv:2311.14589 [hep-th]} \BibitemShut
  {NoStop}%
\bibitem [{\citenamefont {Danielsson}\ \emph {et~al.}(2023)\citenamefont
  {Danielsson}, \citenamefont {Henriksson},\ and\ \citenamefont
  {Panizo}}]{Danielsson:2023aa}%
  \BibitemOpen
  \bibfield  {author} {\bibinfo {author} {\bibfnamefont {U.}~\bibnamefont
  {Danielsson}}, \bibinfo {author} {\bibfnamefont {O.}~\bibnamefont
  {Henriksson}},\ and\ \bibinfo {author} {\bibfnamefont {D.}~\bibnamefont
  {Panizo}},\ }\bibfield  {title} {\bibinfo {title} {{Stringy realization of a
  small and positive cosmological constant in dark bubble cosmology}},\ }\href
  {https://doi.org/10.1103/PhysRevD.107.026020} {\bibfield  {journal} {\bibinfo
   {journal} {Phys. Rev. D}\ }\textbf {\bibinfo {volume} {107}},\ \bibinfo
  {pages} {026020} (\bibinfo {year} {2023})},\ \Eprint
  {https://arxiv.org/abs/2211.10191} {arXiv:2211.10191 [hep-th]} \BibitemShut
  {NoStop}%
\bibitem [{\citenamefont {Ooguri}\ and\ \citenamefont
  {Vafa}(2017)}]{Ooguri:2016pdq}%
  \BibitemOpen
  \bibfield  {author} {\bibinfo {author} {\bibfnamefont {H.}~\bibnamefont
  {Ooguri}}\ and\ \bibinfo {author} {\bibfnamefont {C.}~\bibnamefont {Vafa}},\
  }\bibfield  {title} {\bibinfo {title} {{Non-supersymmetric AdS and the
  Swampland}},\ }\href {https://doi.org/10.4310/ATMP.2017.v21.n7.a8} {\bibfield
   {journal} {\bibinfo  {journal} {Adv. Theor. Math. Phys.}\ }\textbf {\bibinfo
  {volume} {21}},\ \bibinfo {pages} {1787} (\bibinfo {year} {2017})},\ \Eprint
  {https://arxiv.org/abs/1610.01533} {arXiv:1610.01533 [hep-th]} \BibitemShut
  {NoStop}%
\bibitem [{\citenamefont {Danielsson}\ and\ \citenamefont
  {Dibitetto}(2017)}]{Danielsson:2016mtx}%
  \BibitemOpen
  \bibfield  {author} {\bibinfo {author} {\bibfnamefont {U.}~\bibnamefont
  {Danielsson}}\ and\ \bibinfo {author} {\bibfnamefont {G.}~\bibnamefont
  {Dibitetto}},\ }\bibfield  {title} {\bibinfo {title} {{Fate of stringy AdS
  vacua and the weak gravity conjecture}},\ }\href
  {https://doi.org/10.1103/PhysRevD.96.026020} {\bibfield  {journal} {\bibinfo
  {journal} {Phys. Rev. D}\ }\textbf {\bibinfo {volume} {96}},\ \bibinfo
  {pages} {026020} (\bibinfo {year} {2017})},\ \Eprint
  {https://arxiv.org/abs/1611.01395} {arXiv:1611.01395 [hep-th]} \BibitemShut
  {NoStop}%
\bibitem [{\citenamefont {Banerjee}\ \emph
  {et~al.}(2020{\natexlab{a}})\citenamefont {Banerjee}, \citenamefont
  {Danielsson},\ and\ \citenamefont {Giri}}]{Banerjee:2020aa}%
  \BibitemOpen
  \bibfield  {author} {\bibinfo {author} {\bibfnamefont {S.}~\bibnamefont
  {Banerjee}}, \bibinfo {author} {\bibfnamefont {U.}~\bibnamefont
  {Danielsson}},\ and\ \bibinfo {author} {\bibfnamefont {S.}~\bibnamefont
  {Giri}},\ }\bibfield  {title} {\bibinfo {title} {{Bubble needs strings}},\
  }\href {https://doi.org/10.1007/JHEP03(2021)250} {\bibfield  {journal}
  {\bibinfo  {journal} {JHEP}\ }\textbf {\bibinfo {volume} {21}},\ \bibinfo
  {pages} {250}},\ \Eprint {https://arxiv.org/abs/2009.01597} {arXiv:2009.01597
  [hep-th]} \BibitemShut {NoStop}%
\bibitem [{\citenamefont {Banerjee}\ \emph
  {et~al.}(2020{\natexlab{b}})\citenamefont {Banerjee}, \citenamefont
  {Danielsson},\ and\ \citenamefont {Giri}}]{Banerjee:2020ab}%
  \BibitemOpen
  \bibfield  {author} {\bibinfo {author} {\bibfnamefont {S.}~\bibnamefont
  {Banerjee}}, \bibinfo {author} {\bibfnamefont {U.}~\bibnamefont
  {Danielsson}},\ and\ \bibinfo {author} {\bibfnamefont {S.}~\bibnamefont
  {Giri}},\ }\bibfield  {title} {\bibinfo {title} {{Dark bubbles: decorating
  the wall}},\ }\href {https://doi.org/10.1007/JHEP04(2020)085} {\bibfield
  {journal} {\bibinfo  {journal} {JHEP}\ }\textbf {\bibinfo {volume} {04}},\
  \bibinfo {pages} {085}},\ \Eprint {https://arxiv.org/abs/2001.07433}
  {arXiv:2001.07433 [hep-th]} \BibitemShut {NoStop}%
\bibitem [{\citenamefont {Padilla}(2005)}]{Padilla:2004mc}%
  \BibitemOpen
  \bibfield  {author} {\bibinfo {author} {\bibfnamefont {A.}~\bibnamefont
  {Padilla}},\ }\bibfield  {title} {\bibinfo {title} {{Infra-red modification
  of gravity from asymmetric branes}},\ }\href
  {https://doi.org/10.1088/0264-9381/22/6/011} {\bibfield  {journal} {\bibinfo
  {journal} {Class. Quant. Grav.}\ }\textbf {\bibinfo {volume} {22}},\ \bibinfo
  {pages} {1087} (\bibinfo {year} {2005})},\ \Eprint
  {https://arxiv.org/abs/hep-th/0410033} {arXiv:hep-th/0410033} \BibitemShut
  {NoStop}%
\bibitem [{\citenamefont {Banerjee}\ \emph {et~al.}(2021)\citenamefont
  {Banerjee}, \citenamefont {Danielsson},\ and\ \citenamefont
  {Giri}}]{Banerjee:2021ab}%
  \BibitemOpen
  \bibfield  {author} {\bibinfo {author} {\bibfnamefont {S.}~\bibnamefont
  {Banerjee}}, \bibinfo {author} {\bibfnamefont {U.}~\bibnamefont
  {Danielsson}},\ and\ \bibinfo {author} {\bibfnamefont {S.}~\bibnamefont
  {Giri}},\ }\bibfield  {title} {\bibinfo {title} {{Dark bubbles and black
  holes}},\ }\href {https://doi.org/10.1007/JHEP09(2021)158} {\bibfield
  {journal} {\bibinfo  {journal} {JHEP}\ }\textbf {\bibinfo {volume} {09}},\
  \bibinfo {pages} {158}},\ \Eprint {https://arxiv.org/abs/2102.02164}
  {arXiv:2102.02164 [hep-th]} \BibitemShut {NoStop}%
\bibitem [{\citenamefont {Danielsson}\ and\ \citenamefont
  {Van~Hemelryck}(2024)}]{Danielsson:2024frw}%
  \BibitemOpen
  \bibfield  {author} {\bibinfo {author} {\bibfnamefont {U.}~\bibnamefont
  {Danielsson}}\ and\ \bibinfo {author} {\bibfnamefont {V.}~\bibnamefont
  {Van~Hemelryck}},\ }\bibfield  {title} {\bibinfo {title} {{Charged Nariai
  black holes on the dark bubble}},\ }\href
  {https://doi.org/10.1088/1361-6382/ad8f8d} {\bibfield  {journal} {\bibinfo
  {journal} {Class. Quant. Grav.}\ }\textbf {\bibinfo {volume} {41}},\ \bibinfo
  {pages} {245011} (\bibinfo {year} {2024})},\ \Eprint
  {https://arxiv.org/abs/2405.13679} {arXiv:2405.13679 [hep-th]} \BibitemShut
  {NoStop}%
\bibitem [{\citenamefont {Danielsson}\ and\ \citenamefont
  {Giri}(2026{\natexlab{a}})}]{Danielsson:2026a}%
  \BibitemOpen
  \bibfield  {author} {\bibinfo {author} {\bibfnamefont {U.}~\bibnamefont
  {Danielsson}}\ and\ \bibinfo {author} {\bibfnamefont {S.}~\bibnamefont
  {Giri}},\ }\bibfield  {title} {\bibinfo {title} {{Work in progress}},\
  }\href@noop {} {\  (\bibinfo {year} {2026}{\natexlab{a}})}\BibitemShut
  {NoStop}%
\bibitem [{\citenamefont {Banerjee}\ \emph {et~al.}(2023)\citenamefont
  {Banerjee}, \citenamefont {Danielsson},\ and\ \citenamefont
  {Giri}}]{Banerjee:2022ree}%
  \BibitemOpen
  \bibfield  {author} {\bibinfo {author} {\bibfnamefont {S.}~\bibnamefont
  {Banerjee}}, \bibinfo {author} {\bibfnamefont {U.}~\bibnamefont
  {Danielsson}},\ and\ \bibinfo {author} {\bibfnamefont {S.}~\bibnamefont
  {Giri}},\ }\bibfield  {title} {\bibinfo {title} {{Features of a dark energy
  model in string theory}},\ }\href
  {https://doi.org/10.1103/PhysRevD.108.126009} {\bibfield  {journal} {\bibinfo
   {journal} {Phys. Rev. D}\ }\textbf {\bibinfo {volume} {108}},\ \bibinfo
  {pages} {126009} (\bibinfo {year} {2023})},\ \Eprint
  {https://arxiv.org/abs/2212.14004} {arXiv:2212.14004 [hep-th]} \BibitemShut
  {NoStop}%
\bibitem [{\citenamefont {Danielsson}\ and\ \citenamefont
  {Giri}(2026{\natexlab{b}})}]{Danielsson:2025aa}%
  \BibitemOpen
  \bibfield  {author} {\bibinfo {author} {\bibfnamefont {U.}~\bibnamefont
  {Danielsson}}\ and\ \bibinfo {author} {\bibfnamefont {S.}~\bibnamefont
  {Giri}},\ }\bibfield  {title} {\bibinfo {title} {{Weak gravity at micron
  scales from dark bubble cosmology and its cosmological consequences}},\
  }\href {https://doi.org/10.1103/819l-m95m} {\bibfield  {journal} {\bibinfo
  {journal} {Phys. Rev. D}\ }\textbf {\bibinfo {volume} {113}},\ \bibinfo
  {pages} {126010} (\bibinfo {year} {2026}{\natexlab{b}})},\ \Eprint
  {https://arxiv.org/abs/2511.21362} {arXiv:2511.21362 [hep-th]} \BibitemShut
  {NoStop}%
\bibitem [{\citenamefont {Banerjee}\ \emph {et~al.}(2024)\citenamefont
  {Banerjee}, \citenamefont {Danielsson},\ and\ \citenamefont
  {Zemsch}}]{Banerjee:2023uto}%
  \BibitemOpen
  \bibfield  {author} {\bibinfo {author} {\bibfnamefont {S.}~\bibnamefont
  {Banerjee}}, \bibinfo {author} {\bibfnamefont {U.}~\bibnamefont
  {Danielsson}},\ and\ \bibinfo {author} {\bibfnamefont {M.}~\bibnamefont
  {Zemsch}},\ }\bibfield  {title} {\bibinfo {title} {{The dark
  bubbleography}},\ }\href {https://doi.org/10.1007/JHEP02(2024)102} {\bibfield
   {journal} {\bibinfo  {journal} {JHEP}\ }\textbf {\bibinfo {volume} {02}},\
  \bibinfo {pages} {102}},\ \Eprint {https://arxiv.org/abs/2311.16242}
  {arXiv:2311.16242 [hep-th]} \BibitemShut {NoStop}%
\bibitem [{\citenamefont {Compere}\ and\ \citenamefont
  {Marolf}(2008)}]{Compere:2008us}%
  \BibitemOpen
  \bibfield  {author} {\bibinfo {author} {\bibfnamefont {G.}~\bibnamefont
  {Compere}}\ and\ \bibinfo {author} {\bibfnamefont {D.}~\bibnamefont
  {Marolf}},\ }\bibfield  {title} {\bibinfo {title} {{Setting the boundary free
  in AdS/CFT}},\ }\href {https://doi.org/10.1088/0264-9381/25/19/195014}
  {\bibfield  {journal} {\bibinfo  {journal} {Class. Quant. Grav.}\ }\textbf
  {\bibinfo {volume} {25}},\ \bibinfo {pages} {195014} (\bibinfo {year}
  {2008})},\ \Eprint {https://arxiv.org/abs/0805.1902} {arXiv:0805.1902
  [hep-th]} \BibitemShut {NoStop}%
\bibitem [{\citenamefont {Apostolopoulos}\ \emph {et~al.}(2009)\citenamefont
  {Apostolopoulos}, \citenamefont {Siopsis},\ and\ \citenamefont
  {Tetradis}}]{Apostolopoulos:2008ru}%
  \BibitemOpen
  \bibfield  {author} {\bibinfo {author} {\bibfnamefont {P.~S.}\ \bibnamefont
  {Apostolopoulos}}, \bibinfo {author} {\bibfnamefont {G.}~\bibnamefont
  {Siopsis}},\ and\ \bibinfo {author} {\bibfnamefont {N.}~\bibnamefont
  {Tetradis}},\ }\bibfield  {title} {\bibinfo {title} {{Cosmology from an AdS
  Schwarzschild black hole via holography}},\ }\href
  {https://doi.org/10.1103/PhysRevLett.102.151301} {\bibfield  {journal}
  {\bibinfo  {journal} {Phys. Rev. Lett.}\ }\textbf {\bibinfo {volume} {102}},\
  \bibinfo {pages} {151301} (\bibinfo {year} {2009})},\ \Eprint
  {https://arxiv.org/abs/0809.3505} {arXiv:0809.3505 [hep-th]} \BibitemShut
  {NoStop}%
\bibitem [{\citenamefont {Banerjee}\ \emph {et~al.}(2013)\citenamefont
  {Banerjee}, \citenamefont {Bhowmick}, \citenamefont {Sahay},\ and\
  \citenamefont {Siopsis}}]{Banerjee:2012dw}%
  \BibitemOpen
  \bibfield  {author} {\bibinfo {author} {\bibfnamefont {S.}~\bibnamefont
  {Banerjee}}, \bibinfo {author} {\bibfnamefont {S.}~\bibnamefont {Bhowmick}},
  \bibinfo {author} {\bibfnamefont {A.}~\bibnamefont {Sahay}},\ and\ \bibinfo
  {author} {\bibfnamefont {G.}~\bibnamefont {Siopsis}},\ }\bibfield  {title}
  {\bibinfo {title} {{Generalized Holographic Cosmology}},\ }\href
  {https://doi.org/10.1088/0264-9381/30/7/075022} {\bibfield  {journal}
  {\bibinfo  {journal} {Class. Quant. Grav.}\ }\textbf {\bibinfo {volume}
  {30}},\ \bibinfo {pages} {075022} (\bibinfo {year} {2013})},\ \Eprint
  {https://arxiv.org/abs/1207.2983} {arXiv:1207.2983 [hep-th]} \BibitemShut
  {NoStop}%
\bibitem [{\citenamefont {Garriga}\ and\ \citenamefont
  {Tanaka}(2000)}]{Garriga:1999yh}%
  \BibitemOpen
  \bibfield  {author} {\bibinfo {author} {\bibfnamefont {J.}~\bibnamefont
  {Garriga}}\ and\ \bibinfo {author} {\bibfnamefont {T.}~\bibnamefont
  {Tanaka}},\ }\bibfield  {title} {\bibinfo {title} {{Gravity in the brane
  world}},\ }\href {https://doi.org/10.1103/PhysRevLett.84.2778} {\bibfield
  {journal} {\bibinfo  {journal} {Phys. Rev. Lett.}\ }\textbf {\bibinfo
  {volume} {84}},\ \bibinfo {pages} {2778} (\bibinfo {year} {2000})},\ \Eprint
  {https://arxiv.org/abs/hep-th/9911055} {arXiv:hep-th/9911055} \BibitemShut
  {NoStop}%
\bibitem [{\citenamefont {Khoury}(2007)}]{Khoury:2006fg}%
  \BibitemOpen
  \bibfield  {author} {\bibinfo {author} {\bibfnamefont {J.}~\bibnamefont
  {Khoury}},\ }\bibfield  {title} {\bibinfo {title} {{Fading gravity and
  self-inflation}},\ }\href {https://doi.org/10.1103/PhysRevD.76.123513}
  {\bibfield  {journal} {\bibinfo  {journal} {Phys. Rev. D}\ }\textbf {\bibinfo
  {volume} {76}},\ \bibinfo {pages} {123513} (\bibinfo {year} {2007})},\
  \Eprint {https://arxiv.org/abs/hep-th/0612052} {arXiv:hep-th/0612052}
  \BibitemShut {NoStop}%
\bibitem [{\citenamefont {Danielsson}\ \emph {et~al.}(2021)\citenamefont
  {Danielsson}, \citenamefont {Panizo}, \citenamefont {Tielemans},\ and\
  \citenamefont {Van~Riet}}]{Danielsson:2021aa}%
  \BibitemOpen
  \bibfield  {author} {\bibinfo {author} {\bibfnamefont {U.~H.}\ \bibnamefont
  {Danielsson}}, \bibinfo {author} {\bibfnamefont {D.}~\bibnamefont {Panizo}},
  \bibinfo {author} {\bibfnamefont {R.}~\bibnamefont {Tielemans}},\ and\
  \bibinfo {author} {\bibfnamefont {T.}~\bibnamefont {Van~Riet}},\ }\bibfield
  {title} {\bibinfo {title} {{Higher-dimensional view on quantum cosmology}},\
  }\href {https://doi.org/10.1103/PhysRevD.104.086015} {\bibfield  {journal}
  {\bibinfo  {journal} {Phys. Rev. D}\ }\textbf {\bibinfo {volume} {104}},\
  \bibinfo {pages} {086015} (\bibinfo {year} {2021})},\ \Eprint
  {https://arxiv.org/abs/2105.03253} {arXiv:2105.03253 [hep-th]} \BibitemShut
  {NoStop}%
\bibitem [{\citenamefont {Danielsson}\ \emph
  {et~al.}(2026{\natexlab{a}})\citenamefont {Danielsson}, \citenamefont {Giri},
  \citenamefont {Mohan},\ and\ \citenamefont {Thorlacius}}]{Danielsson:2026b}%
  \BibitemOpen
  \bibfield  {author} {\bibinfo {author} {\bibfnamefont {U.}~\bibnamefont
  {Danielsson}}, \bibinfo {author} {\bibfnamefont {S.}~\bibnamefont {Giri}},
  \bibinfo {author} {\bibfnamefont {V.}~\bibnamefont {Mohan}},\ and\ \bibinfo
  {author} {\bibfnamefont {L.}~\bibnamefont {Thorlacius}},\ }\bibfield  {title}
  {\bibinfo {title} {{Work in progress}},\ }\href@noop {} {\  (\bibinfo {year}
  {2026}{\natexlab{a}})}\BibitemShut {NoStop}%
\bibitem [{\citenamefont {Bekenstein}(1974)}]{Bekenstein:1974jk}%
  \BibitemOpen
  \bibfield  {author} {\bibinfo {author} {\bibfnamefont {J.~D.}\ \bibnamefont
  {Bekenstein}},\ }\bibfield  {title} {\bibinfo {title} {{The quantum mass
  spectrum of the Kerr black hole}},\ }\href
  {https://doi.org/10.1007/BF02762768} {\bibfield  {journal} {\bibinfo
  {journal} {Lett. Nuovo Cim.}\ }\textbf {\bibinfo {volume} {11}},\ \bibinfo
  {pages} {467} (\bibinfo {year} {1974})}\BibitemShut {NoStop}%
\bibitem [{\citenamefont {Mukhanov}(1986)}]{Mukhanov:1986me}%
  \BibitemOpen
  \bibfield  {author} {\bibinfo {author} {\bibfnamefont {V.~F.}\ \bibnamefont
  {Mukhanov}},\ }\bibfield  {title} {\bibinfo {title} {{Are black holes
  quantized?}},\ }\href@noop {} {\bibfield  {journal} {\bibinfo  {journal}
  {JETP Lett.}\ }\textbf {\bibinfo {volume} {44}},\ \bibinfo {pages} {63}
  (\bibinfo {year} {1986})}\BibitemShut {NoStop}%
\bibitem [{\citenamefont {Bekenstein}\ and\ \citenamefont
  {Mukhanov}(1995)}]{Bekenstein:1995ju}%
  \BibitemOpen
  \bibfield  {author} {\bibinfo {author} {\bibfnamefont {J.~D.}\ \bibnamefont
  {Bekenstein}}\ and\ \bibinfo {author} {\bibfnamefont {V.~F.}\ \bibnamefont
  {Mukhanov}},\ }\bibfield  {title} {\bibinfo {title} {{Spectroscopy of the
  quantum black hole}},\ }\href {https://doi.org/10.1016/0370-2693(95)01148-J}
  {\bibfield  {journal} {\bibinfo  {journal} {Phys. Lett. B}\ }\textbf
  {\bibinfo {volume} {360}},\ \bibinfo {pages} {7} (\bibinfo {year} {1995})},\
  \Eprint {https://arxiv.org/abs/gr-qc/9505012} {arXiv:gr-qc/9505012}
  \BibitemShut {NoStop}%
\bibitem [{\citenamefont {Basile}\ \emph {et~al.}(2024)\citenamefont {Basile},
  \citenamefont {Danielsson}, \citenamefont {Giri},\ and\ \citenamefont
  {Panizo}}]{Basile:2023aa}%
  \BibitemOpen
  \bibfield  {author} {\bibinfo {author} {\bibfnamefont {I.}~\bibnamefont
  {Basile}}, \bibinfo {author} {\bibfnamefont {U.}~\bibnamefont {Danielsson}},
  \bibinfo {author} {\bibfnamefont {S.}~\bibnamefont {Giri}},\ and\ \bibinfo
  {author} {\bibfnamefont {D.}~\bibnamefont {Panizo}},\ }\bibfield  {title}
  {\bibinfo {title} {{Shedding light on dark bubble cosmology}},\ }\href
  {https://doi.org/10.1007/JHEP02(2024)112} {\bibfield  {journal} {\bibinfo
  {journal} {JHEP}\ }\textbf {\bibinfo {volume} {02}},\ \bibinfo {pages}
  {112}},\ \Eprint {https://arxiv.org/abs/2310.15032} {arXiv:2310.15032
  [hep-th]} \BibitemShut {NoStop}%
\bibitem [{\citenamefont {Danielsson}\ \emph
  {et~al.}(2026{\natexlab{b}})\citenamefont {Danielsson}, \citenamefont
  {Panizo},\ and\ \citenamefont {Van~Hemelryck}}]{Danielsson:2026qch}%
  \BibitemOpen
  \bibfield  {author} {\bibinfo {author} {\bibfnamefont {U.}~\bibnamefont
  {Danielsson}}, \bibinfo {author} {\bibfnamefont {D.}~\bibnamefont {Panizo}},\
  and\ \bibinfo {author} {\bibfnamefont {V.}~\bibnamefont {Van~Hemelryck}},\
  }\bibfield  {title} {\bibinfo {title} {{Self-gravitating electromagnetic
  waves in the dark bubble model}},\ }\href@noop {} {\  (\bibinfo {year}
  {2026}{\natexlab{b}})},\ \Eprint {https://arxiv.org/abs/2606.16547}
  {arXiv:2606.16547 [hep-th]} \BibitemShut {NoStop}%
\end{thebibliography}%

\end{document}